\magnification=1200
\baselineskip=18truept
\input epsf

%preprint or not
\def\preprint{Y}
%draft or not
\def\draftversion{N}

\def\chiral{{\bf\rm C}}
\def\wilson{{\bf\rm B}}
\def\ham{{\bf\rm H}}
\def\mbham{{\cal H}}
\def\wblU{{\!\langle}}
\def\wbrU{{\rangle^{\rm WB}_U}}
\def\wbl1{{{}^{\rm WB}_{\ \ 1}\!\langle}}

\def\cap{\hsize=4.5in}

\if \draftversion Y


\fi

% Figures
\def\figure#1#2#3{\if \preprint Y \midinsert \epsfxsize=#3truein
\centerline{\epsffile{figure_#1_eps}} \halign{##\hfill\quad
&\vtop{\parindent=0pt \hsize=5.5in \strut## \strut}\cr {\bf Figure
#1}&#2 \cr} \endinsert \fi}

\def\figureb#1#2{\if \preprint N \midinsert \epsfxsize=#2truein
\centerline{\epsffile{figure_#1_eps}} \halign{##\hfill\quad
&\vtop{\parindent=0pt \hsize=5.5in \strut## \strut}\cr \cr \cr
\cr \cr \cr  {\bf Figure #1} \cr} \endinsert \fi}

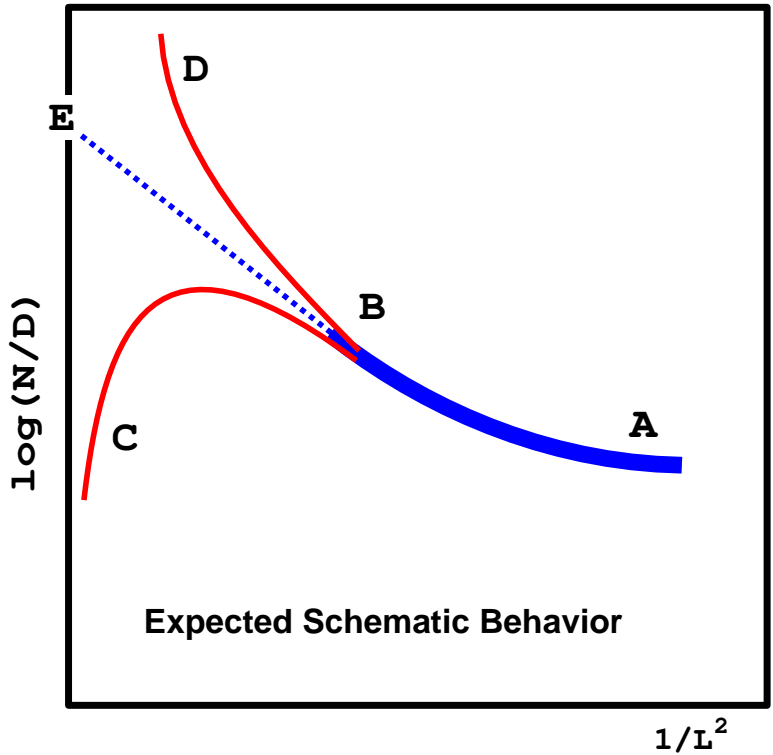
\def\captionone {\cap Effect of Thirring coupling. }
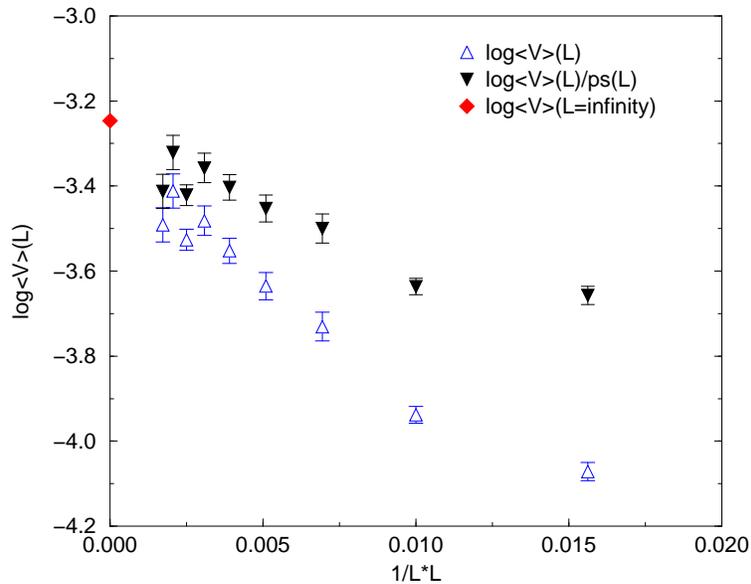
\def\captiontwo {\cap Data obtained with gauge averaging. Convergence
improves by inclusion of the  point-split factor $ps$.}
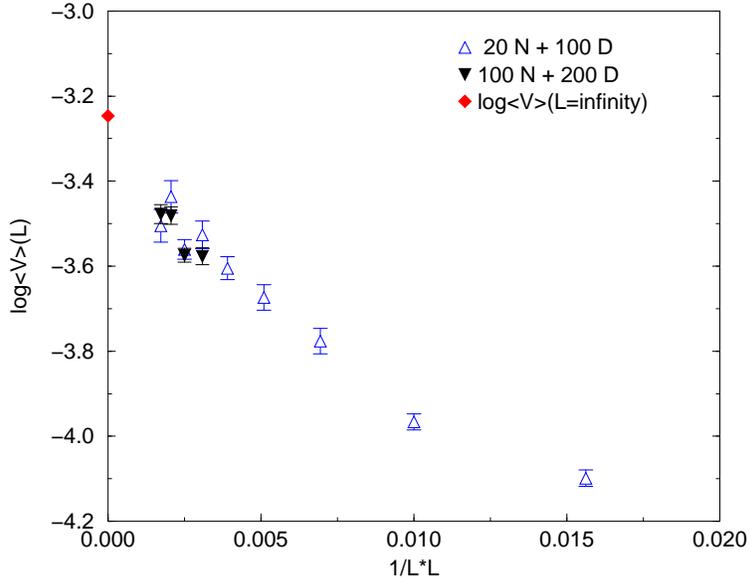
\def\captionthree {\cap Data obtained using the absolute value and no
gauge averaging. For $L=18,20,22,24$ higher statistics points
are added. }
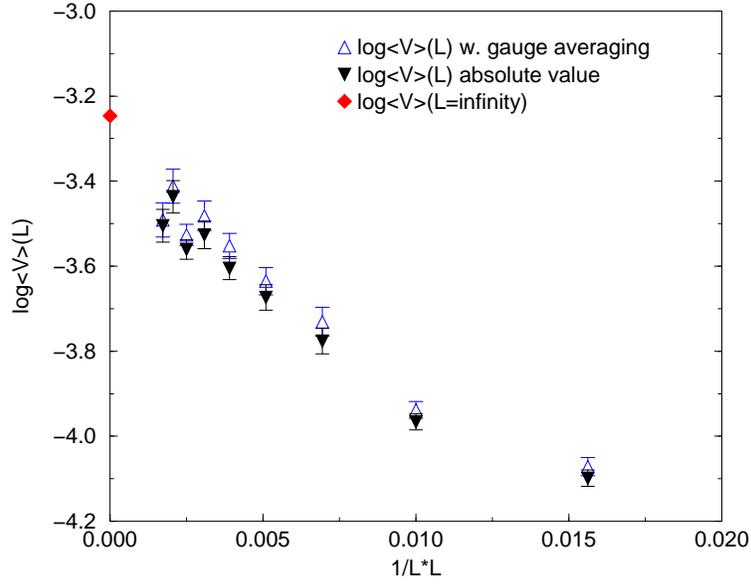
\def\captionfour {\cap A comparison of data obtained with gauge averaging
to data obtained without for the same sample of orbits. }
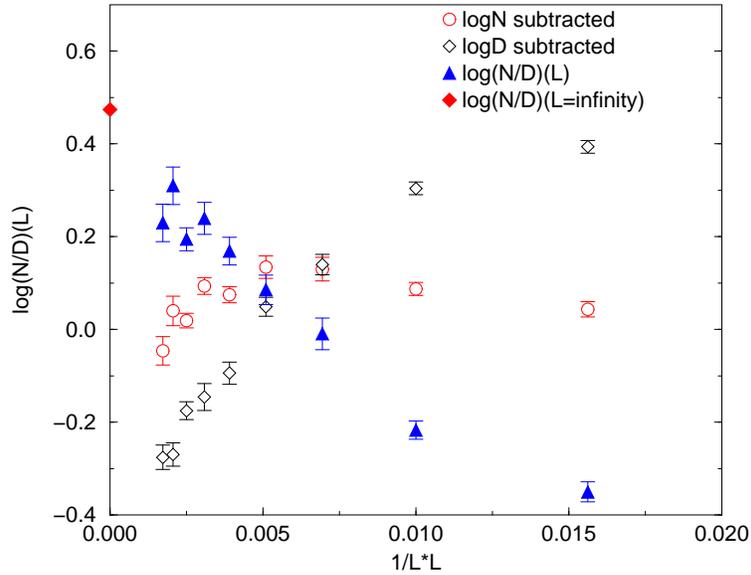
\def\captionfive {\cap $\log N$ and $\log D$ after subtracting the
quadratic divergence are small numbers. The logarithm of $N/D$ 
is also shown. The numbers were obtained with gauge averaging. }

\line{\hfill RU-97-28}
\line{\hfill UW/PT-97-10}
\line{\hfill DOE/ER-40561-322-INT97-00-169}
\line{\hfill KUNS-1443  HE(TH)97/06}
\vskip 2truecm
\centerline{\bf Monte Carlo evaluation of a 
fermion number violating observable in 2D.}

\vskip 1truecm
\centerline{Yoshio Kikukawa${}^{a}$\footnote{*}{Permanent Address: 
Department of Physics, Kyoto University, Kyoto 606-01, Japan.
}, Rajamani Narayanan${}^{b}$
and Herbert Neuberger${}^{a}$}
\vskip .5truecm

\centerline {${}^a$ Department of Physics and Astronomy}
\centerline {Rutgers University, Piscataway, NJ 08855-0849}
\centerline {${}^b$ Institute for Nuclear Theory, Box 351550}
\centerline {University of Washington, Seattle, WA 98195-1550}
\vskip 1.5truecm

\centerline{\bf Abstract}
\vskip 0.75truecm
We describe in some detail a computer evaluation of a 
't Hooft vertex in a two dimensional model using the overlap.
The computer result agrees with the known exact continuum value,
and in this sense our work is a first successful fully dynamical
simulation of a chiral gauge theory on the lattice. 
We add some new data to numbers obtained earlier  
and provide a selfcontained description
which should make it easy for others to reproduce and follow
up on our work. 

\vfill
\eject

\centerline
{\bf 1. Introduction}
\medskip
Unlike for QCD, a communal base of numerical
experience about simulating chiral gauge theories 
on the computer is lacking. Even on what should be 
done in principle there is no consensus; in 
practice, handling the generically complex action
is likely to present a formidable problem by 
itself.
        One approach for regularizing chiral gauge
theories, the overlap [1], has been recently subjected
to a full dynamical test for a simple model [2]. The
result was positive and therefore a more detailed
description of what was done in practice has 
become relevant. Previous publications emphasized
theory and were brief on numerics. Here our 
objective is to collect all the technical details 
needed to carry out our test, so that the reader 
could reproduce and follow up on our numerical work 
without having to  sift through several papers. 
The reader would be helped
by QCD experience only to a limited extent; because
of the presumed lack of familiarity with the
overlap in practice our presentation
will be quite explicit. 

\bigskip
\centerline{\bf 2. Continuum features.}
\medskip
Our particular model is a chiral abelian gauge theory in two Euclidean
dimensions. The theory is defined on a torus of sides $l_1=t$, $l_2 =l$.
Matter consists of four left-handed Weyl fermions, $\chi_f$, of charge
one and one right-handed Weyl fermion of charge two, $\psi$. 
The action is:
$$
S={1\over 4e_0^2} \int d^2 x F^2_{\mu\nu}
-\sum_{f=1}^4\int d^2 x \bar\chi_f \sigma_\mu 
(\partial_\mu+iA_\mu )\chi_f
-\int d^2 x \bar\psi \sigma^*_\mu (\partial_\mu+2iA_\mu )\psi ,\eqno{(2.1)}$$
where $\sigma_1=1$, $\sigma_2=i$ and $\mu =1,2$. The $U(1)$ gauge symmetry
is anomaly free by $2^2 = 1^2 +1^2 +1^2 +1^2 $. The boundary conditions are:
$$\eqalign{
\chi_f(x + l_\mu \hat\mu)= e^{2\pi i b^f_\mu}\chi_f(x ) \cr
\psi(x+ l_\mu \hat\mu)=\psi(x )\cr
F_{\gamma\nu}(x+ l_\mu \hat\mu)=F_{\gamma\nu} (x)}\eqno{(2.2)}$$
for $\mu =1,2$. $\hat\mu$ is a unit vector in the $\mu$ direction. The
$\bar\chi_f$ and $\bar\psi$ fields obey complex conjugate boundary conditions.
The $b_\mu^f$ are given by:
$$b^1_1=0;\ \ \ b^2_1=0;\ \ \ b^3_1={1\over 2};\ \ \ b^4_1={1\over 2};
\ \ \ \ \
b^1_2=0;\ \ \ b^2_2={1\over 2};\ \ \ b^3_2=0;\ \ \ b^4_2={1\over 2}.
\eqno{(2.3)}$$

In the $t=l=\infty$ limit the following properties are relevant to the present
work:
\item{$\bullet$} The physical Hilbert space can be written as a tensor product 
of two spaces which are not connected by the evolution operator. One of the
spaces carries a representation of a massless conformal theory and the other 
describes a massive field theory. If we replaced $\psi$-matter by four
{\it right} moving Weyl fermions of charge one the theory would be vectorial.
Again, a similar factorization would hold and the massive factor would
be isomorphic to the massive factor in our original chiral model.
\item{$\bullet$} The massless sector consists of six left moving  Majorana
Weyl fermions forming a sextet under the global $SU(4)$ acting on the
$f$-index of the $\chi_f$'s. These particles are noninteracting. One can
choose interpolating fields for these particles which are neutral objects
local in the original fields:
$$
\rho_{f_1 f_2} = - \rho_{f_2 f_1}= {{\pi^{3\over 2} e^{-\gamma}}\over
e_0} [\chi_{f_1} \chi_{f_2} \bar\psi - {1\over 2} \epsilon^{f_1 f_2 
f_3 f_4 } \bar\chi_{f_3 }
\bar\chi_{f_4 } \psi ]. \eqno{(2.4)}$$
The prefactor is chosen so that $\rho$ becomes a canonical field at large
distances. $\gamma$ is Euler's constant. 
The mixing on the RHS is possible because
fermion number is violated
in quanta of two by instanton effects. 
The exact low energy effective Lagrangian  of the model, written
in terms of the $\rho$-fields, is
$$
{\cal L} ={1\over 2} \sum_{f_1 > f_2 } \rho_{f_1 f_2 } \sigma\cdot
\partial \rho_{f_1 f_2 }.\eqno{(2.5)}$$
\item{$\bullet$} Among the terms on the RHS we have a term proportional 
to a 't Hooft 
vertex, $V(x)$, which we choose to define as:
$$
V(x)={{\pi^2}\over
{e_0^4}} \chi_1(x) \chi_2(x) \chi_3(x) \chi_4(x)
\bar\psi(x) (\sigma\cdot\partial )\bar\psi(x).
\eqno{(2.6)}$$
This operator has a nonzero expectation value:
$$
\langle V \rangle_\infty ={{e^{4\gamma}}\over 
{4\pi^3}}\approx 0.081~.\eqno{(2.7)}$$
On the finite $t\times l$ lattice one has instead:
$$\langle V\rangle_{t\times l} = {64\pi \over (t m_\gamma)^4} 
\exp \Bigl[ -{4\pi\over  t m_\gamma}
\coth \left ( {1\over 2}lm_\gamma \right ) \Bigr] 
e^{4F( t m_\gamma)-8H( t m_\gamma , {t \over l} )},
\eqno{(2.8)}$$
where $m_\gamma^2 = {{4e_0^2}\over \pi}$ and the functions $F(\xi )$ and 
$H(\xi , \tau )$ are defined below:
$$\eqalign{
F(\xi )& =\sum_{n>0} \Bigl
[ {1\over n} - {1\over \sqrt{n^2+(\xi /2\pi)^2}}\Bigr],\cr
H(\xi ,\tau)& =\sum_{n>0}{1\over \sqrt{n^2+(\xi /2\pi)^2}}
{1\over e^{\tau\sqrt{(2\pi n)^2+\xi^2 }}-1}.\cr}\eqno{(2.9)}$$
We shall need $\langle V\rangle_{t\times l}$ 
at $t=l={3\over m_\gamma}$ and there
its value is 0.0389, substantially smaller than the value at infinite volume.
\vskip .15in
The objective of our numerical work is to 
reproduce this last continuum number on
the lattice. Namely, we wish to evaluate the ratio of formal path integrals:
$$
{{\int \prod_{f=1}^4 [ d\bar\chi^f 
d \chi^f ] [d\bar\psi d\psi ] \prod_{\mu=1}^2
{[ dA_\mu ]} ~ e^S ~ V}\over
{\int \prod_{f=1}^4 [ d\bar\chi^f 
d \chi^f ] [d\bar\psi d\psi ] \prod_{\mu=1}^2
[ dA_\mu ] ~e^S ~}}\equiv \langle V \rangle.\eqno{(2.10)}$$
The path integrals are rendered finite by replacing the torus by a toroidal
square lattice with spacing $a$. 
To preserve invariance under rotations by ninety degrees 
the two sides of the lattice are made equal. 
One wishes to show that the correct
continuum number emerges in the limit $a\to 0$.

The following decomposition of $A_\mu$ simplifies the 
path integrals in the continuum:
$$
A_1=\partial_2 \phi + 
{2\pi\over l} h_1 + i g^{-1} \partial_1 g ,\ \ \ \ \ 
A_2= -{{2\pi q}\over l^2} x_1 - \partial_1 \phi + 
{2\pi\over l} h_2 + i g^{-1} \partial_2 g .\eqno{(2.11)}
$$
Above, $g$ is a periodic complex valued
function on the torus with $|g(x)|\equiv 1$, representing the gauge degree of
freedom. $\phi$ is a real periodic function on the torus and has no zero mode:
$\int d^2 x \phi \equiv 0$. $h_\mu$ are two real constants restricted
to the intervals $(-1/2, 1/2]$ which can be thought of as parameterizing
two Polyakov loops on the torus. $q$ is an integer identifying the topological
class. 
The pure gauge
part of $S$ depends only on $\phi$ and $q$:
$$
{1\over 4e_0^2} \int d^2 x F^2_{\mu\nu} = {1\over 2 e_0^2} 
\int d^2 x (\partial^2 \phi )^2 + {{2\pi^2 q^2}\over 
{(e_0 l)^2}}\equiv S_0 +s_q .\eqno{(2.12)}$$
The gauge measure $[ dA_\mu ] $ is replaced 
by $\sum_q [ dg ]  d^2 h [ d\phi ] $. 
Because the number of fermionic zero modes is determined
by $q$, only 
the $q=1$ term contributes in the 
sum over $q$ in the numerator of (2.10) and only the $q=0$ term contributes 
to the denominator there. Thus, the sum over $q$ collapses to a single term.
Gauge invariance implies that the integral over $g$ can be ignored. Only
the fermionic part of the integrand depends on $h_\mu$.

\bigskip
\centerline{\bf 3. Lattice formulation}
\medskip

On the lattice the ``measurement'' of $<V>$ proceeds by two separate
Monte Carlo simulations: one for the numerator and the other
for the denominator in (2.10). To the numerator only gauge fields
which induce exactly one fermionic zero mode for each
charge one fermion and exactly two zero modes 
for the doubly charged fermion contribute. 
The denominator is entirely determined
by configurations with strictly no fermion zero modes. 
In principle, given a gauge configuration, one needs to
perform some calculations to count the zero modes. 
However, by electing to work with a noncompact pure gauge action
it becomes possible also on the lattice to predict
the number of zero modes without a calculation simply
by setting $q$. It is still up to the 
fermionic part of the system to ``agree''
with our prescribed $q$. We monitor for ``disagreements''
where the number of fermionic zero modes on the lattice
is not what it should be for the given $q$. With our
choice of parameters such disagreements are so rare
that we never encounter them during our simulations.
The main gain from this is that we can set $q$ by hand
and our result does not suffer from the statistical
noise inherent in simulating the prefactor $e^{-s_q}$
from (2.12).

Our objective is to obtain the expectation
value of the 't Hooft vertex in a fixed, rather small, physical
volume. This is implemented by setting the bare coupling, $ e_0$, 
to $ 1.5{\sqrt{\pi}\over L}$ when we work on a finite 
$L\times L$ lattice. The continuum limit is taken by
increasing $L$. Note that the parameter $m_\gamma$
was introduced as a bare parameter below equation (2.8).
{\it A priori}, there is no guarantee that the lack of
renormalization of the bare coupling in the continuum should hold
also on the lattice. We assume that it does, and if
our final result agrees with the continuum
value we can say that the numerics 
are consistent with this assumption.

\smallskip
\centerline{\sl 3A. Bosonic variables.}
\smallskip

We use the same notation
as in the continuum for $\phi , g, h, q$. The pure gauge action is

$$S_0 +s_q\equiv 
{1\over 2 e_0^2} 
\sum_ x (\Delta \phi )^2 + {{2\pi^2 q^2}\over 
{(e_0 L )^2}},\eqno{(3.1)}$$
where 
$$\Delta \phi (x) = \sum_\mu (\phi(x+\hat\mu) +\phi (x-\hat\mu)) -4\phi (x),
~~~~~~~~~\phi (x+L\hat\mu )\equiv \phi (x) .\eqno{(3.2)}$$
 
All site index components $x_\mu$ are between $0$ and $L-1$
and if they appear outside this range in any
of our equations one has to use the appropriate
boundary conditions to bring them back within range. 
The field $\phi$ is constrained by $\sum_x \phi (x) = 0$.
In Fourier space
this constraint is trivially implemented. 
The pure gauge integration measure is taken as:
$$
\sum_{q\in Z} e^{-s_q}
\int d\mu \cdots \equiv 
\sum_{q\in Z} e^{-s_q}
\int \prod dg(x) 
\int_{-1/2}^{1/2} d^2 h 
\int_{\sum_x \phi (x) = 0} \prod_x d\phi (x) e^{-S_0 (\phi)} 
\cdots
\eqno{(3.3)}
$$
and the pure gauge partition function is
$$Z = \sum_{q\in Z} e^{-s_q} {\cal N};
\ \ \ \ \ 
{\cal N} = \int d\mu .\eqno{(3.4)}$$
Both Monte Carlo simulations carried out to get $<V>$ will
calculate the expectation value of appropriate ``fermion
observables'' in an ensemble governed by $d\mu$. In other
words, both the numerator and the denominator of (2.10)
are divided by ${\cal N}$. The ``fermion observables'' 
are functions of $g, h, \phi$ that represent lattice
versions of what in continuum is viewed as the result
of carrying out the Grassmann integral in a fixed gauge
background. With $q$ fixed the pure gauge action depends only on
$\phi$, just like in the continuum. The ``fermion observable''
depends also on $g$, in addition to $h_\mu$. Hence, the
$g$ integral cannot be dropped.

We generate gauge fields using the pure gauge action and treat
the fermion integrals in a fixed gauge field background as 
observables in the pure gauge theory. In our Monte Carlo
implementation
there is a difference between the numerator ond the denominator.
The measure for generating the variables $h$ is flat in (3.3).
We shall keep it this way for the numerator. In the denominator
however, we replace the flat measure for $h$ by a measure
including the known $h$-dependent factor from the
continuum fermionic contribution. This factor is extracted back from
the fermionic observable (the equivalent of the chiral determinant),
so the answer is unchanged. We are thus 
exploiting the fact that the
dependence on $h$ factorizes from the fermionic determinant
in the continuum and from the pure gauge 
action both in the continuum
and on the lattice. This trick significantly reduces the statistical
error of the Monte Carlo simulation ``measuring'' the denominator.

For the numerator the $h_\mu$ are generated randomly 
in the interval $[-1/2,1/2]$.
For the denominator the $h_\mu $ are generated by a weight
proportional to: 
$$|\theta(2h_1,2h_2)|^2 = \Bigg|\sum_{n=-\infty}^\infty e^{-\pi(n+2h_2)^2 + 
2i\pi n 2h_1} \Bigg|^2 .
\eqno{(3.5)}$$

The other gauge variables are simulated in a standard way:
$S_0$ defined in (3.1) and (3.2)
can be written in momentum space as
$$S_0 = {1\over 2\pi} \sum_{k_0,k_1=0}^{L-1} \tilde\phi(k) \tilde\phi(-k)
\Bigl[ 4 \sin^2 ({\pi k_0\over L}) + 4 \sin^2 ({\pi k_1\over L}) \Bigr ]^2 ,
\eqno{(3.6)}$$
where
$$\phi(x) = {e_0\over L\sqrt{\pi}}
\sum_{k_0,k_1=0}^{L-1} \tilde\phi (k) e^{{2\pi i\over L}k\cdot x};
\ \ \ \ \ \tilde\phi(-k) = \tilde\phi^*(k);\ \ \ \ \tilde\phi(0)=0 
\eqno{(3.7)}$$
and $-k_\mu \equiv {\rm mod}(L-k_\mu,L)$. 
If $k \ne -k$, we generate two random numbers, $u_1,u_2$, normally
distributed according to
$\int du \exp \Biggl [ -{u^2 \over \pi}
\Bigl[ 4 \sin^2 ({\pi k_0\over L}) + 4 \sin^2
({\pi k_1\over L}) \Bigr ]^2
\Biggr ]$
and set $\tilde\phi(k) = u_1 + i u_2$.
If $k = -k$, we generate one random number, $u$, normally
distributed according to
$\int du \exp \Biggl [- {u^2 \over 2\pi}
\Bigl[ 4 \sin^2 ({\pi k_0\over L}) + 4 \sin^2 ({\pi k_1\over L}) \Bigr ]^2
\Biggr ]$
and set $\tilde\phi(k) = u$.
We then use (3.7) to obtain 
several independent $\phi(x)$ configurations.

For the denominator we can therefore replace $d\mu$ in (3.3) by
$$\eqalign{
\int \prod dg(x) &
\int_{-1/2}^{1/2} d^2 h 
\int_{-1/2}^{1/2} d^2 r 
\prod_\mu \Bigl[ \delta_{h_\mu, {r_\mu\over 2}} +
\delta_{h_\mu, {1\over 2} (r_\mu -{r_\mu\over |r_\mu|})} \Bigr]
{|\theta(r_1,r_2)|^2\over z_t} \cr
&
\int_{\sum_x \phi (x) = 0} \prod_x d\phi (x) e^{-S_0 (\phi)} 
{z_t\over |\theta(r_1,r_2)|^2}
\cdots \cr}
\eqno{(3.8)}$$
where
$z_t= 4\int_{-1/2}^{1/2} d^2 r {|\theta(r_1,r_2)|^2}$
and the last factor is thought of as part of the observable.
In this manner we obtain $h_\mu$ in the range $[-(1/2),(1/2)]$
distributed according to (3.7).
Note that $q$ is presumed fixed already.
For the numerator, there are no $r, z_t$ variables and the factors
containing them are absent. 
Finally, the gauge variables $g(x)$ are U(1) gauge transformations 
randomly generated at each site $x$. Each gauge configuration we generate is
statistically independent of the previous ones if we assume 
a high quality random number generator. We used RANLUX at luxury level 4.
\footnote{*}{Subtract-and-borrow random number generator proposed by
Marsaglia and Zaman, implemented by F. James with the name 
RCARRY in 1991, and later improved by Martin L{\" u}scher 
in 1993 to produce "Luxury Pseudorandom Numbers". 
Fortran 77 coded by F. James, 1993 [3]. 
}

The fermions do not depend on the noncompact fields directly. 
They only depend on ordinary, compact, link variables. 
Thus, if one wished to
replace the noncompact pure fermionic action by
a compact one the fermionic part of our code would remain 
unchanged. The connection between the variables 
$g, h, \phi, q$ we 
generate as described above 
and the link fields transmitted to the fermions is given by:

$$
U_1(x) = 
\cases{ 
g(x) e^{2\pi i h_1\over l}
e^{[\phi(x) - \phi(x-\hat 1)]} 
e^{{2\pi iq\over L} x_2} g^*(x+\hat 1)
& if $x_1=L-1$ \cr
g(x) e^{2\pi i h_1\over l}
e^{[\phi(x) - \phi(x-\hat 1)]}
g^*(x+\hat 1)
& otherwise \cr}
$$
$$\eqno{(3.9)}$$
$$
U_2(x) = 
g(x) e^{2\pi i h_2\over l}
e^{[\phi(x-\hat 2) - \phi(x)]}
e^{-{2\pi iq\over L^2} x_1} g^*(x+\hat 2).
$$
Note that for $q\ne 0$ we
take care of the needed twist 
at the boundary in the first line of (3.9). The
noncompact character of the pure gauge action is
reflected in that the mapping $\{ g,h,\phi , q\}\rightarrow 
\{U_\mu \}$ is many to one.

The right handed fermions live on the lattice the gauge fields
are defined on. The parallel transporters between neighboring sites
for these doubly charged fermions are $U_\mu (x)^2$. Up to gauge
transformations the $U_\mu (x)$  are quite close
to the identity. The $U^2$ fields will be a little farther
from the identity.
The deviations of $i(1-U^2)$ from $2A$ are cutoff effects of some
typical magnitude. Similar cutoff effects also exist for the 
singly charged fermions and there is no 
numerical reason to try to make
them smaller than the ones for the doubly charged fermions. 
We can thus put the charge one fermions
on a twice as coarse lattice, thereby saving some computer time without
increasing the ultraviolet cutoff effects. 

\smallskip
\centerline{\sl 3B. Fermionic variables.}
\smallskip

The definition of the fermionic observables is in terms of
fermionic creation and annihilation operators associated with
the continuum Grassmann variables. Unlike their
continuum counterparts, these operators are Dirac spinors (not
Weyl spinors) thus ending up with twice as many components.
The charge 2 $\psi(x)$ field is represented by $b(x;\alpha)$
where $\alpha =1,2$ is the extra spinor index. The charge
1 $\chi_f (x)$ fields are represented by $a_f (y;\alpha)$.
The sites $y$ are on the coarse lattice and their integer
components
range sequentially from $0$ to ${L\over 2} -1$. 
$L$ is always even. Whenever
a site index is outside its fundamental range one has to 
use the following boundary conditions to bring it back 
within range:
$$b(x + L\hat\mu_x ;\alpha) = b(x_\mu;\alpha),\ \ \ \ 
a_f(y+{L\over 2} \hat\mu_y ;\alpha) = e^{2\pi i b^f_\mu} a_f (y;\alpha),
\eqno{(3.10)}$$
where $b^f_\mu$ are given by (2.3).
The operators $a,b$  obey canonical anti-commutation relation and
act on the usual Fock space. $\mu_x$ and $\mu_y$ are elementary
links on the fine and coarse lattices respectively.

For the
numerator we need to discretize (2.6). To this end
we introduce the operators $\eta_{f_1 f_2 }(y)$ below:
$$\eta_{f_1 f_2 }(y) = a_{f_1} (y;2) a_{f_2}(y;2) 
b^\dagger(2y;2) .\eqno{(3.11)}$$
The site
argument of $b$ is relative to the fine lattice. All operators
in (3.11) reside at the same physical site. 
The spinorial indices are chosen to match the continuum
Lorentz transformation properties of the original fields.
Thus, the $\eta$'s are lefthanded and $U(1)$ neutral.
With the help of the $\eta$'s we can produce a gauge invariant
point split discretization of $V$. The $\eta$ fields
also appear in the definition of the $\rho$ field in (2.4).

The central object of the overlap is a Hamiltonian that acts
on the Fock space irreducibly 
representing the complete algebra
of anti-commutators. The Hamiltonian is bilinear and
separable by flavor. These two properties represent
the bilinearity and separability of the continuum action.
The Hamiltonian (with implicit 
summations over spin and site indices) is given by:
$$\mbham = b^\dagger \ham^b (U^b ) b 
-\sum_{f=1}^4 a_f^\dagger \ham^a ( U^a ) a_f .\eqno{(3.12)}$$

The single particle Hamiltonians appearing in (3.12) are given below:

$$\ham^{a,b} (U^{a,b} ) = 
\pmatrix{ \wilson(U^{a,b} ) - m & \chiral(U^{a,b}) \cr
\chiral^\dagger(U^{a,b} ) & -\wilson(U^{a,b} ) + m \cr },
\eqno{(3.13)}$$
$$
\chiral(z, z^\prime;U) =
{1\over 2}
\sum_{\mu=1}^2 \sigma_\mu
\Bigl [
\delta_{z^\prime ,z +\hat\mu_z } (U_\mu(z)) -
\delta_{z,z^\prime +\hat\mu_z } (U^*_\mu(z^\prime ))
\Bigr] ,\eqno{(3.14)}
$$
$$
\wilson(z, z^\prime ;U) =
{1\over 2} \sum_{\mu=1}^4
\Bigl [
2 \delta_{zz^\prime} -
\delta_{z^\prime ,z+\hat\mu_z } (U_\mu(z)) -
\delta_{z,z^\prime +\hat\mu_z } (U^*_\mu(z^\prime ))
\Bigr] .\eqno{(3.15)}
$$
(3.13) displays the dependence on spin indices in matrix
format while (3.14-15) define the site dependence explicitly.
When using (3.13-15) in (3.12), equation (3.10) must
be employed to obtain the correct form of $\mbham$,
with site index coordinates restricted to $[0,L-1]$ for $b$
and to $[0,L/2 -1]$ for $a_f$. 
$U^{a,b}$ are the link variables as seen by the
appropriate fermions. In our case the $a$ fermions
reside on the coarse lattice and the $U^a$ link
variable live on the links of that coarse lattice.
The $U^b$ link variables are on the original lattice and 
are chosen as identical to the link variables $U$ introduced
in (3.9). The $U^a$ link variables are locally defined in terms
of the $U$'s as follows:
The parallel transporter
between $y$ and $y+\hat\mu_y$ on the coarse lattice, $ U^a_\mu (y)$, 
is given by:
$$ 
\eqalign { U^a_\mu (y) = & {W \over |W|} ,\cr 
W = & 
U_\mu (2y) U_\mu (2y+\hat\mu_x ) + 
U_\nu (2y) U_\mu (2y+\hat\nu_x ) 
U_\mu (2y+\hat\nu_x +\hat\mu_x ) 
U^*_\nu (2y+2\hat \mu_x )\cr 
& + U^*_\nu (2y-\hat\nu_x ) 
U_\mu (2y-\hat\nu_x ) 
U_\mu (2y-\hat\nu_x +\hat\mu_x ) 
U_\nu (2y+2\hat\mu_x -\hat\nu_x ),\ \ \ \
\nu\ne\mu .\cr }
\eqno{(3.16)}
$$

To obtain
a proper description for chiral fermions one needs to 
pick $0 < m < 1$. In practice, we used $m=.5$. This
is a value empirically tuned to minimize potentially UV divergent
effects due to a dynamically generated 
Thirring coupling.

The fermionic observables are matrix elements of fermionic
operators made out of $a,b$ in a way corresponding to the
continuum expression in terms of Grassmann variables. 
The matrix elements are taken between two different states:
The ``ket'' is the ground state of $\mbham$, $ | 0  \wbrU$, with 
the phase fixed in a manner which is dependent on the 
topology:  

$$
\wbl1 0 |
(\tilde a_1^\dagger (2) \tilde a_1 (1)
\tilde b (1)\tilde b^\dagger (2)|0 \wbrU \eqno{(3.17)}
$$ 
is real and positive if $q=0$ and
$$
\sum_{{}^{f_1 ,f_2 ,f_3 ,f_4 =1}}^4 \sum_{\mu=1}^2 \sum_y 
~ \epsilon_{f_1 f_2 f_3 f_4 }~ 
\sigma_\mu ~\wbl1 0|
(\tilde a_1^\dagger (2) \tilde a_1 (1)
\tilde b (1)\tilde b^\dagger (2))~
\eta_{f_1 f_2 }(y)~  \eta_{f_3 f_4 }(y+\hat\mu_y ) | 0 \wbrU 
\eqno{(3.18)}
$$
is real and positive if $q=1$. The phase of the state $\wbl1 0|$
is immaterial since it does not depend on $U$. 
The sum over $y$ is over all sites on the
coarse lattice. $\tilde a_1 (\alpha) = \sum_y a_1(y;\alpha)$ is the 
operator associated with a single particle zero momentum
state for the $q=1$ fermion that has periodic boundary conditions
in both directions and the sum on $y$ is over all points on the
coarse lattice.
Similarly, $\tilde b (\alpha ) = \sum_x b(x;\alpha )$ is the 
operator associated with a single particle zero momentum
state for the $q=2$ fermion while the sum on $x$ is over all points
on the fine lattice. 
These operators create states that carry no momentum. The 
combinations are also charge neutral. 
All insertions are best thought of as
acting to the left and defining a $U$-independent reference state
used to fix the phase of the ground state of the $U$-dependent
system. The reference state is a Lorentz scalar and carries no 
$U(1)$ charge.

The ``bra'' state in the matrix element representing the result
of the Grassmann integration, $ \wblU 0 |$,  is 
$U$-independent and given by the ground state of an ultralocal
Hamiltonian, similar to (3.12) only that the single particle
Hamiltonians (3.13)  are diagonal  having only mass terms
with the parameter $m$ there replaced by a negative number.
The phase choice for the ``bra'' is immaterial as long as it keeps
the state $U$-independent.

With these definitions in place we finally can write down 
the lattice formula for (2.10) in the overlap formalism on an $L\times L$
lattice as:
$$ \langle V \rangle = {{2 \pi^2}\over 
{(e_0 L)^4}} e^{-{2\pi^2\over (e_0 L )^2}}
{N\over D} = 0.024204 {N\over D} ,\eqno{(3.19)}$$ 
$$
N = {{L^2} \over 24} 
\sum_{{}^{f_1 , f_2 , f_3 , f_4 =1}}^4 \sum_{\mu}^2 \sum_y 
\sigma_{\mu_y} ~\epsilon_{f_1 f_2 f_3 f_4 }~
\int   {  d\mu\over {\cal N}} ~ 
\wblU 0  | \eta_{f_1 f_2 }(y) \eta_{f_3 f_4 }(y+\hat\mu_y ) | 0  \wbrU
,\eqno{(3.20)}
$$
$$ D = \int {d\mu\over {\cal N}}  ~
\wblU 0  | 0  \wbrU .
\eqno{(3.21)}
$$

The sum over $y$ in (3.20) is over all points on the coarse lattice.

Expressions (3.20) and (3.21) are still not in an explicit
enough form to transcribe directly into a program. First we
need to rewrite (3.20) and (3.21) in first quantized
forms. To this end, we diagonalize the single particle
Hamiltonians (3.13):
$$
\ham O = O \Lambda .\eqno{(3.22)}$$
The $\Lambda$ are diagonal matrices and the real
elements along the diagonal are in decreasing order top to bottom.
We work with five Hamiltonians: four are $L^2 /2 \times L^2 /2 $ matrices
and differ by the boundary conditions on the unit charged fermions.
The fifth is an $2L^2 \times 2L^2$ matrix representing the doubly
charged fermion. 

We also define some gauge field independent reference matrices $O_0$: These
will be used in implementing the Wigner-Brillouin phase definition.
We shall need a realization of the bra state in (3.17) 
and (3.18) after being
acted on by the $\tilde a_1$ and $\tilde b$ operators.

$$
\ham_0 O_0  = O_0 \Lambda_0  .\eqno{(3.23)}$$
$\ham_0$ is one of the five free Hamiltonian obtained
by setting the gauge links to unity. The boundary conditions
are maintained. 
For a trivial gauge configuration one can diagonalize the
matrices by Fourier transform. Except for fermions which
are periodic in both directions the ordering of $\Lambda_0$
is the same as above. For periodic fermions
there are zero momentum eigenstates and the action of 
the $\tilde a_1$ and $\tilde b$ operators is implemented 
by switching the position of the positive
eigenvalue corresponding to zero total momentum 
with the
position of the negative eigenvalue corresponding 
to zero momentum. This induces
a switch of two columns in $O_0 $

It is convenient to partition the $O$ matrices into the
following pattern:

$$
O_a^f = \pmatrix {
O^f_{LL}(U^a) & O^f_{LR}(U^a) \cr 
O^f_{RL}(U^a) & O^f_{RR}(U^a)\cr},\ \ \ \ f=1,2,3,4 ;~~~~~~O_b = \pmatrix {
O_{LL}(U^b) & O_{LR}(U^b) \cr
O_{RL}(U^b) & O_{RR}(U^b) \cr} .\eqno{(3.24)} $$

The submatrices have half as many rows as the original $O$ matrices.
These rows are indexed by a site index, running over $L^2$ values
for $O_b$ (defined on the fine lattice) and over $L^2 /4$ values
for $O_a^f$ (defined on the coarse lattice). The first columns
of $O$ contain the eigenvectors corresponding to positive eigenvalues
of $\ham$. This set of columns defines the horizontal partition
of the $O$ matrices.  (Exactly zero eigenvalues of $\ham$ are possible
but are non-generic.) The $O_0$ 
reference matrices are partitioned into square submatrices
$O_{0LL}, O_{0LR}, O_{0RL}, O_{0RR}$ similarly to (3.24). 

We define a configuration to have zero lattice topology (in practice,
these are configurations with $q=0$) when all submatrices are square.
The lattice topological number will be unity (in practice, these
are configurations with $q=1$) when all $O_{LL}^f$ are rectangular
with $L^2 /4 +1$ columns and $O_{LL}$ is rectangular with $L^2 +2$
columns. When the lattice topological number is unity we
associate with the rectangular matrices the following site
dependent square matrices:

$$\eqalign{&
{O^f_{LL}(U^a;y ) = \pmatrix{O^f_{LL}(U^a)\cr v^f_{RL}(U^a;y ) \cr}},\cr
& {O_{RR}(U^b;x_y , \mu ) = \pmatrix{O_{RR}(U^b),& v(x_y ) & 
v(x_y +\hat\mu_y ) \cr}},\cr
&  {O_{LR}(U^b;0,0 ) = \pmatrix{O_{LR}(U^b),& 0, & 0\cr }}.\cr}
\eqno{(3.25)}$$
The site $y$ is on the coarse lattice. $x_y \equiv 2y$ 
identifies a site on the fine
lattice that also resides on the coarse lattice. $\hat\mu_y$ connects
two sites of the coarse lattice. $v^f_{RL}(U^a;y)$ is the $y^{\rm th}$ 
row of $O^f_{RL}(U^a)$. $v(x_y)$ is a column vector of 
length $L^2$ that is zero everywhere except
at the site $x_y$ and ``0''
denotes a zero column vector. 
The $O^f_{LL}(U^a;y)$ are $(L^2 /4 +1)\times (L^2 /4 +1)$
matrices. $O_{RR}(U^b;y,\mu)$ and $ O_{LR}(U^b;0,0 )$
are $L^2 \times L^2$ matrices.
The site dependence of the square matrices defined in (3.24) corresponds
to the sites of the $\eta$ operators in (3.20).

We need some extra rectangular reference matrices 
for the case of unit lattice topology:
$$
O_{0LL}^{f} (0)= \pmatrix{O_{0LL}^f ,& 0\cr},~~~~
O_{0RL}^f (y) = \pmatrix{O_{0RL}^f, & v(y) \cr}.\eqno{(3.26)}$$

Now we can write down the single particle representations of the
matrix elements in (3.20) and (3.21).
The overlap appearing in (3.21) has the following explicit form:
$$\wblU 0  | 0  \wbrU = \prod_f \det O^f_{LL}
(U^a) \det O_{RR}(U^b) e^{i\phi(U)}
\eqno{(3.27)}$$
with
$$e^{i\phi(U)} =
\Bigl [
{\det {\rm WB}_{RR}(U^b) \over |\det {\rm WB}_{RR}(U^b)|}\Bigr ]
\prod_f \Bigl [
{\det {\rm WB}^f_{LL}(U^a) \over |\det {\rm WB}^f_{LL}(U^a)|}\Bigr ].
\eqno{(3.28)}$$
In (3.28),
$$\eqalign{
{\rm WB}_{RR}(U^b)& = \bigl[O_{RR}(U^b)\bigr]^\dagger O_{0RR}
+\bigl[O_{LR}(U^b)\bigr]^\dagger O_{0LR} ,\cr
{\rm WB}^f_{LL}(U^a)& = \bigl[O^f_{LL}(U^a)\bigr]^\dagger O^f_{0LL}
+\bigl[O^f_{RL}(U^a)\bigr]^\dagger O^f_{0RL} .\cr}
\eqno{(3.29)}$$

The overlap appearing in (3.20) becomes (no sum over $f_i$),
$$\eqalign{
\wblU 0  | &\eta_{f_1 f_2 }(y) \eta_{f_3 f_4 }(y+\hat\mu) | 0  \wbrU
= \epsilon_{f_1 f_2 f_3 f_4 } 
\det O^{f_1}_{LL}(U^a;y) \det O^{f_2}_{LL}(U^a;y) \cr
  & \det O^{f_3}_{LL}(U^a;y+\hat\mu_y ) \det O^{f_4}_{LL}(U^a;y+\hat\mu_y ) 
  \det O_{RR}(U^b;x_y ,\mu) e^{i\phi(U)}\cr}\eqno{(3.30)}$$
with
$$
 e^{i\phi(U)}={{\rm wb}(U)\over |{\rm wb}(U)|}.\eqno{(3.31)}$$
In (3.31), 
$$\eqalign{
{\rm wb}(U) = & 
\sum_{{}^{f_1 ,f_2 ,f_3 ,f_4 =1}}^4 \sum_{\mu=1}^2 \sum_y 
~ [ \epsilon_{f_1 f_2 f_3 f_4 } ]^2 ~ 
\sigma_\mu 
\det {\rm WB}^{f_1}_{LL}(U^a;y) \det {\rm WB}^{f_2}_{LL}(U^a;y)\cr 
 & \det {\rm WB}^{f_3}_{LL}(U^a;y+\hat\mu_y 
) \det {\rm WB}^{f_4}_{LL}(U^a;y+\hat\mu_y ) 
\det {\rm WB}_{RR}(U^a;x_y ,\mu) .\cr}\eqno{(3.32)}$$
The matrices appearing in (3.32) are given by
$$\eqalign{&
{\rm WB}^f_{LL}(U^a;y) = [ O^f_{LL}(U^a) ]^\dagger O^f_{0LL}( 0) + 
[ O^f_{RL}(U^a) ]^\dagger  O^f_{0RL}(y)  ,\cr
& {\rm WB}_{RR}(U^b;x_y ,\mu) = 
[ O_{RR}(U^b ; x_y ,\mu )]^\dagger  O_{0RR}  + 
[ O_{LR}(U^b ; 0,0 )]^\dagger  O_{0LR} .\cr} \eqno{(3.33)}$$

\smallskip
\centerline{\sl 3C.  The Monte Carlo procedure.}
\smallskip

We have seen that the bosonic variables are integrated over by Monte Carlo
while the fermionic variables are ``integrated'' out ``exactly''.  The
bosonic integration has two stages, consisting of two Monte Carlo processes,
one running within the other. The outer process (for either $q=0$ or $q=1$)
averages over gauge orbits,
generating the fields $\phi (x) $ and $h_\mu$. The links on the coarse
and fine lattice are then defined up to a gauge transformation. Setting
$g(x)\equiv 1$ in (3.9) this gives us a representative of an orbit. Loosely,
we can refer to this as the ``Landau'' gauge representative, since in the
continuum, setting $g$ to unity in (2.11) would imply $\partial_\mu A_\mu =0$.
We then compute the needed determinants (3.27-33), depending on the topological
sector we are in. 

The internal Monte Carlo procedure averages
over gauge transformations after a ``Landau'' configuration
(an orbit) has been selected and the fermionic observables have
been calculated. For each $x$ we randomly draw a unimodular
complex number $g(x)$.
When the links are replaced by
gauge transformed links (see (3.9)) the dependence 
of the the Hamiltonians $\ham^{a,b}$ in (3.13) on the
gauge transformation $g$ can be absorbed into  
a unitary matrix $G^{a,b}$ which conjugates the 
``Landau'' gauge  $\ham^{a,b}$.  $G^{a}$ ( $G^{b}$ )  
is a diagonal $L^2 /2 \times L^2 /2 $ ($2 L^2 \times 2 L^2 $) 
matrix whose rows and columns are indexed by
sites and spin components. The entries on the diagonal
are the appropriate $g$ for the site.
Thus the Hamiltonians
after the gauge transformation are expressed in terms
of the ``Landau'' gauge Hamiltonians as:
$$
\ham^{a,b} ({\rm gauge ~transformed ~links})~ =~ 
[G^{a,b}] ~\ham^{a,b}(
{\rm Landau ~gauge~ links})~
[ G^{a,b} ]^\dagger .\eqno{(3.34)}
$$ 
The effect of the conjugation  in 
(3.34) is  to multiply the ``Landau''
gauge  $O^f_a$ and $O_b$ from the left by $G^{a,b}$.
The submatrices (3.24) and (3.25) also get multiplied
from the left by appropriate diagonal submatrices of
$G^{a,b}$. The determinant factors in (3.27) and (3.30) 
change by gauge factors given by
the product of the five determinants of these
diagonal submatrices. Because of gauge invariance, the
dependence on the site $y$ in (3.30) cancels out from
its gauge factor and both gauge factors turn out to
be given by $\prod_x g(x) [ \prod_y g(y) ]^4$ 
(we remind the reader that the $x$'s 
are sites on the fine lattice
and the $y$'s are sites on the embedded coarse lattice).

The diagonal matrices implementing the gauge transformations
on the matrices $O$ get caught in between the matrices multiplied
in (3.29) and (3.33) and the related determinants in
(3.28) and (3.32) have to be recomputed. Since no rediagonalization
of the Hamiltonians is necessary, and since the changes in (3.27) 
and (3.30) are simple, the total amount of computation needed
to account for link updates along gauge orbits is smaller
than for link updates changing gauge orbit. 

Every orbit we generate is completely independent statistically
(assuming an ideal random number generator) from every other
orbit. Similarly, all gauge transformations we generate are mutually
independent. The absence of correlations implies that statistical
error analysis is relatively simpler here than in standard
Monte Carlo. Of course, since our observables are non-local, one
needs to look at their distributions to ascertain that they are
reasonable and that our numerical approach hasn't been rendered
hopeless by large fluctuations. For sizes $L=8-24$ the fluctuations
are manageable and we can make quite accurate statistical
determinations. The ability to control the fluctuations
is helped by the adjustment of the coupling constant with $L$ which
keeps the physical volume finite. 

For a given gauge configuration
the number of floating point operations is quite high, $\sim 10^{11}$ 
for the highest $L$'s. The evaluations of the gauge averages of 
the imaginary parts involve some cancelations which
are roundoff sensitive. Thus, it is necessary to use
double precision floating point arithmetic throughout (except
in the generation of the random numbers).

\bigskip
\centerline{\bf 4. Numerical results.}
\medskip

We evaluated the quantities $N$ and $D$ ((3.20-21)) for all even
$L$'s between 8 and 24. For a single gauge configuration, the
evaluation of a $D$-contribution is faster than the evaluation
of an $N$-contribution. This is countered somewhat by smaller
fluctuations for $N$. We aimed at comparable accuracies in $N$ 
and $D$ and ended up spending ten times as much computer time 
on evaluating $N$ than on calculating $D$ for 
the larger $L$'s. For a fixed gauge 
configuration and asymptotic $L$, $N$ is an $L^8$ computation 
while $D$ is an $L^6$ computation.

It was determined that reasonable values could be obtained for $D$
with 100 orbits plus 100 gauge transforms per orbit. For $N$
20 orbits with 40 gauge transforms sufficed. This yielded
relative errors (single standard deviation) in $N$ and $D$ of a few
percent. 

Numerically, $N$ and $D$ are dominated by ultraviolet effects.
When $L$ is large we expect the following asymptotics:
$$\eqalign{
\log (N) &= n_0 L^2 + n_1 \log L + n_2 + 
{{n_3 \log L }\over L^2} + {{n_4 }\over L^2} + \cdots \cr
\log (D) &= d_0 L^2 + d_1 \log L + d_2 + {{d_3 \log L}
\over L^2} + {{d_4} \over L^2 } + \cdots\cr}
\eqno{(4.1)}$$ 
$n_0$ and $d_0$ are evaluated from the noninteracting theory:
$$n_0 = d_0 = 0.242586~.\eqno{(4.2)}$$

A regularized version 
cannot reproduce (2.1)
exactly in the continuum limit because 
an exactly marginal (in the renormalization group sense)
coupling exists and obeys all symmetries. This is the
``Thirring'' coupling:
$$
g_{Th} ( \sum_f \bar\chi_f \chi_f )\bar\psi\psi .\eqno{(4.3)}$$  
$g_{Th}$ can have either sign. Depending on it we can have $n_1 > d_1$ 
or $n_1 < d_1$.

Generically, by tuning a free parameter in the regularization 
one can adjust the approach to continuum to have $n_1 = d_1$. 
In our case this is approximatively achieved by
choosing $m=.5$ (see paragraph after (3.16)). Of course, the
tuning cannot be exact, so we will expect $N/D$ to either diverge or
go to zero as $L$ becomes truly large. But, as long as $|n_1-d_1|\log L$
is numerically negligible $\log(N/D)$ is dominated by $n_2 - d_2$.
The first subleading correction is of order $1/L^2$ since there
are no dimension three irrelevant operators to contribute to the
action or the operator. The exact result quoted in section 2
is essentially $n_2 -d _2$ for the case $n_1 = d_1$. 

Schematically, we would expect the following 
behavior of the ratio as a function of $1/L^2$:

\figure{1}{\captionone}{4.0}

The data would follow either ABC or ABD, depending on the sign of $g_{Th}$.
The better we tuned $m$ the closer will B by to E, the
exact result for $g_{Th}=0$. This is only the simplest
scenario for what happens around B. Since various terms make
comparable contributions the curves may get more complicated
there. 

The $1/L^2$ corrections that come from the operator (rather than from the
action) can be evaluated at  $g_{Th}=0$ in the continuum. This yields
a correction factor (point-split factor) of:

$$ ps = \exp  \{  -[ \int_0^\infty {{d\tau} \over \tau}
e^{-{3\over L} ( \tau +{1\over \tau }) }  + 
1.15443133 + 2\log {3\over L}] \}.
\eqno{(4.4)} $$  
The factor $ps$ can be used to improve convergence. The number 
1.15443133 above is two times Euler's constant and $3/L$ comes 
from $2 e_0 / \sqrt{\pi }  $. We use the complete formula for $ps$,
including all subleading terms in $1/L^2 $. 
\figure{2}{\captiontwo}{4.3}

Figure 2 contains our results for the logarithm of the
vertex, $\log <V> $. 
The exact
continuum limit is also 
shown ( $\log <V>(L=\infty ) = -3.247$ ). 
Quite clearly, we see convergence
towards the exact continuum number for $L$ between 8 and 18. $ps$
indeed ``improves'' the observable. The data for $L$ between 20 and
24 show a certain amount of instability. It is conceivable that
this reflects region B in our schematic figure. We cannot say
anything conclusive without better statistics and additional
results for higher $L$'s.
\figure{3}{\captionthree}{4.3}
Most of the computation time is spent on gauge averaging. By 
discarding gauge averaging we can increase statistics. 
A simple gauge invariant way to avoid gauge averaging is
to replace the integrands in (3.20) and (3.21) by their absolute
values. The corresponding quantities in the continuum 
are positive. In figure 3 we show data obtained
using the absolute values exploiting the same runs as in figure
1. We also add four more data points in the range $L=18$ to 24
with higher statistics: $N$ is evaluated from 100 orbits and $D$ from
200. We see that the higher statistics, while supportive of some
structure around $L=20$, still do not lead us to a definite
confirmation of the qualitative scenario that would relate it 
to a Thirring term.
\figure{4}{\captionfour}{4.3}
\figure{5}{\captionfive}{4.3}
To be sure that using the absolute value rather than the
gauge averaged results is not a source of contamination
as far as the possibility of a structure around $L=20$ is concerned
we compare in figure 4 the two (absolute value and gauge averaged)
vertices for the entire set of available orbits. 
We see that the difference is quite insignificant numerically 
and, therefore, how gauge invariance is recovered is
unimportant.

In figure 5 we display $N$ and $D$ individually. 
We subtract the large numbers $n_0 L^2 = d_0 L^2$ from the logarithms
of $N$ and $D$ and plot the results together with
the logarithm of the ratio ($\log (N/D) = 0.474$
in the continuum limit). We see that the bulk of $D$ 
and $N$ is given by the free field terms we subtracted, and that
the remainder consists of numbers of order unity. 

\bigskip
\centerline{\bf 5. Summary.}
\medskip

Although the tuning of the Thirring coupling is not fully understood
numerically, overall, our simulation confirms the validity of the
overlap regularization method of the model in (2.1). 
More work would be needed to understand the possible role
of a dynamically generated $g_{Th}$. 
Other two dimensional models 
can be tested by similar methods.
Going to higher dimensions
would require some algorithmical improvements, as 
the simulations presented here took a considerable amount of time on
machines of various types. 

Based on our work, we express our hope that the problem of simulating
chiral gauge theories has been turned from one of principle
to one of practicality.

\bigskip
\bigskip
\centerline{\bf  Acknowledgments.}
\medskip

The research of R. N. was supported in part by the 
DOE under grant \# DE-FG03-96ER40956 and \# DE-FG06-90ER40561 and that of
H. N. was supported in part by the DOE under grant \#
DE-FG05-96ER40559.

\bigskip
\bigskip
\bigskip
\centerline{\bf  References.}
\medskip
\item{[1]} R. Narayanan, H. Neuberger, 
Nucl. Phys. B443 (1995) 305.
\item{[2]} R. Narayanan, H. Neuberger,
Phys. Lett. B, to appear, hep-lat/9609031.  
\item{} Y. Kikukawa, R. Narayanan, 
H. Neuberger, Phys. Lett. B, to appear, \hfill 
\item{} hep-lat/9701007. \hfill
\item{[3]} M. L{\" u}scher, Computer Physics Communications  79 (1994) 100. 
\item {}   F. James, Computer Physics Communications 79 (1994) 111.

\vfill\eject\end